# Strong Emergence Arising from Weak Emergence

Thomas Schmickl

Artificial Life Lab, Institute of Biology, Department of Zoology, University of Graz, Austria

thomas.schmickl@uni-graz.at

Abstract – Predictions of emergent phenomena, appearing on the macroscopic layer of a complex system, can fail if they are made by a microscopic model. This study demonstrates and analyses this claim on a well-known complex system, Conway's Game of Life. Straightforward macroscopic mean-field models are easily capable of predicting such emergent properties after they are fitted to simulation data in an after-the-fact way. Thus, these predictions are macro-to-macro only. However, a micro-to-macro model significantly fails to predict correctly, as does the obvious mesoscopic modeling approach. This suggests that some macroscopic system properties in a complex dynamic system should be interpreted as examples of phenomena (properties) arising from "strong emergence", due to the lack of ability to build a consistent micro-to-macro model, that could explain these phenomena in a before-the-fact way. The root cause for this inability to predict this in a micro-to-macro way is identified as the pattern formation process, a phenomenon that is usually classified as being of "weak emergence". Ultimately, this suggests that it may be in principle impossible to discriminate between such distinct categories of "weak" and "strong" emergence, as phenomena of both types can be part of the very same feedback loop that mainly governs the system's dynamics.

## 1. INTRODUCTION

The tendency to exhibit emergent properties is a key characteristic of complex systems [1]. Emergence has been studied and discussed in various scientific domains, such as biology [2], ecology [3], neurology [4], economics [5], social sciences [6][7], linguistics [8], and philosophy [9]. In all complex systems that are studied in these domains, emergent properties arise on the macroscopic system layer, caused by feedback loops that reside on a microscopic layer. For example, often microscopic and localized feedback loops induce self-organizational processes, which produce patterns that then appear on the macroscopic system layer.

There is significant debate about the existence and the relevance of the phenomenon of emergence in complex systems, which led to the proposal of discriminating between two categories of emergence: "weak emergence" and "strong emergence" [10]. However, this distinction between the two variants of emergence did not settle the debate. On the one hand, it prompted questions about this distinction itself, while on the other hand debates about the mere existence of "strong emergence" persisted. For example, Mark Bedau, states "Although strong emergence is logically possible, it is uncomfortable like magic" [10], while Peter Corning concludes "Emergence […] is neither a mystical concept nor is it a threat to reductionist science" [11].

For the sake of simplicity only a short description of the used terms is given here to define how the terms "emergent phenomena", "weak emergence" and "strong emergence" are used in the study at hand:

Emergent phenomena: Novel phenomena properties of a complex system, that are often surprising at first sight. They arise on a higher (macro) system level as a non-trivial consequence of the microscopic mechanics, that operate on a lower system level. The non-triviality arises usually from non-linear component interactions, which create feedback loops that are associated with specific timing coefficients (e.g., delays of causal effects) within complex systems.

Weak emergence: Phenomena or system properties that may surprise the observer at first sight, but which are easily explainable with micro-to-macro causation afterwards. By thinking hard, or with an appropriate model, such phenomena should also be able to be predicted without having observed them first. This basically means that such phenomena on the macroscopic level can be predicted by applying a model that is purely based on the known microscopic mechanisms of a system. However, predicting such phenomena with a macroscopic model, that is parametrized from macroscopic observations, would not suffice to identify an emergent phenomenon as being "weakly emergent", as such a model would not explain how and why the microscopic mechanisms cause and govern the observed macroscopic phenomenon.

Strong emergence: These are emergent phenomena or system properties which cannot be explained (post-hoc) and predicted (a priori) by a consistent model (chain of micro-to-macro causation) from the underlying microscopic mechanisms. This is for example the case when the emergent phenomenon on the macroscopic system level feeds back to the microscopic components and modulates their behavior. This way a closed loop of micro-macro-micro causation arises. For example, there might be a strong micro-to-macro effect that is easily explained by the microscopic mechanism and thus only weakly emergent, if it is emergent at all. Even subtle feedback from macro-to-micro might affect exactly these microscopic weak emergent processes, as in nonlinear interaction systems, even weak feedbacks can have strong ultimate effects and drive the system towards alternative states, e.g., through phase-transitions. Thus, such feedbacks can impair the ability to produce a reliable micro-to-macro model of the system, turning this inability into a characteristic of strong emergence.

One might first think, that claiming that an observed phenomenon or system property is of strong emergence, is an exceptional claim. Thus, one would infer that such a claim would need to be supported by exceptionally strong evidence. However, in contrast to this first intuition, it seems to me that the exact opposite approach is required. The study presented here aims at demonstrating that the burden of proof in fact lies on the claim of weak emergence for interpreting an emergent property:

In my opinion, the *really* exceptional claim, that needs to be proven, is to consider an emergent phenomenon to be of weak emergence, as this basically means, that one *can* explain the observed emergent phenomenon purely from the known microscopic mechanisms. In consequence, if an emergent property is observed, one should consider it to be strong emergence at first and then try to falsify this hypothesis by coming up with a micro-to-macro model. This micro-to-macro model should explain the phenomenon and allow predictions in sufficient quality. The harder this model building task gets, the more likely the observed phenomenon might be an example of strong emergence, as for real strong emergence it will be impossible to find such a sufficient micro-to-macro model. Ultimately, this way of interpreting strong emergence means that weakly emergent phenomena can be components in the explanation of strongly emergent phenomena and vice versa.

The fact that complex systems exhibit emergent macroscopic features unexplainable from a purely microscopic model has been demonstrated for properties of the Ising model [12] in a decidability study of a system composed of an Ising model in combination with a Turing machine [13].

The study presented here aims at demonstrating that strongly emergent properties can be observed already in complex systems with very simple mechanisms, such as Conway's Game of Life (GoL) [14], which is still, decades after its discovery, subject to comprehensive research in various communities [15]. The seminal article by Marc Bedau studies the GoL for emergence and describes several properties of this system to be phenomena of the weak emergence type: pattern formation in general and specifically the growth of a cell population starting from the small "pentomino configuration" in an otherwise empty world [10].

These phenomena and system properties can all be easily explained by studying and modeling the simple microscopic rules of the GoL. However, does the simple GoL also produce phenomena that are cases of strong emergence?

The hypothesis that is investigated in this article is that specific system properties, that seem to be very trivial at first sight, might in fact be the most non-trivial system properties and examples for strong emergence: The global population dynamics (changes of population densities over time) and the long-term non-zero population density (LTNPD) towards which the system converges, are both observable in the GoL.

The main hypothesis of this study is that these properties, which are statistical properties of the system that one could easily observe and then analyze from performing forward simulations, are unpredictable from analyzing the microscopic rules without simulating the system. In order to investigate this hypothesis, the focal research questions of the study presented here are:

1. Are there emergent macroscopic properties in the GoL, that are usually not considered to be emergent?
2. Can we come up with a micro-to-macro model to explain these system properties by "weak emergence"?
3. Can we come up with a mesoscopic model to explain these system properties by these microscopic mechanisms, still following a bottom-up approach, thus being purely informed a-priori by the microscopic mechanisms?
4. Are these phenomena easily predictable by a macroscopic post-hoc model?
5. How do these properties and modeling approaches work out for other cellular automata that are not exactly the GoL but similarly acting on the microscopic level?

## 2. Conway's Game of Life

In the following, a very brief description of the GoL is given to explain those aspects that are necessary to understand the arguments and procedures described in this study.

The GoL acts in a finite grid-type (lattice) "world" of width $X$ and height $Y$. In the study presented here the dimensions are $X = Y = 201$ cells. Each cell at any position $(x, y)$ at any time step $t$ is either in state $S(x, y, t) = 0$ (dead) or in state $S(x, y, t) = 1$ (alive). Every time step, the cells update their state, according to the rules depicted in Table 1. This update is synchronous. Thus, all cells update to their new state $S(x, y, t + 1)$ based on the states of their neighboring cells and their own state in their previous time step $t$. For this update, each cell considers its own state and the number of "living" cells in the eight adjacent cells within its Moore neighborhood [16]. This individual "neighborhood count" is expressed by $N(x, y, t)$. For neighborhood assessment the world is wrapped in a toroid (donut-type) topology to avoid edge effects, thus the world is finite concerning the number of cells, but it has no outer edges. The exact update rules of the GoL are given in Table 1, basically, all configurations yield a "dead" cell state except:

- A new cell is born if a dead cell had 3 living neighbors (green frame in Table 1).
- A cell stays alive with 2 or 3 living neighbor cells (blue frames in Table 1).

Table 1: Ruleset of the Game of Life

| if the focal cell *at (x,y)* is dead at time step *t*: | | | | | | | | | |
|---|---|---|---|---|---|---|---|---|---|
| N(x, y, t) | 0 | 1 | 2 | 3 | 4 | 5 | 6 | 7 | 8 |
| S(x, y, t) | 0 | 0 | 0 | 0 | 0 | 0 | 0 | 0 | 0 |
| S(x, y, t+1) | 0 | 0 | 0 | 1 | 0 | 0 | 0 | 0 | 0 |
| if the focal cell *at (x,y)* is alive at time step *t*: | | | | | | | | | |
| N(x, y, t) | 0 | 1 | 2 | 3 | 4 | 5 | 6 | 7 | 8 |
| S(x, y, t) | 1 | 1 | 1 | 1 | 1 | 1 | 1 | 1 | 1 |
| S(x, y, t+1) | 0 | 0 | 1 | 1 | 0 | 0 | 0 | 0 | 0 |

After all cells have successfully updated their states, the number of living cells for each time step can be calculated by $N_{living}(t) = \sum_{x=1}^{X}\sum_{y=1}^{Y} S(x, y, t)$ and the fraction of living cells (population density) in the total population can be calculated as $\Omega(t) = N_{living}(t)/(X \cdot Y)$.

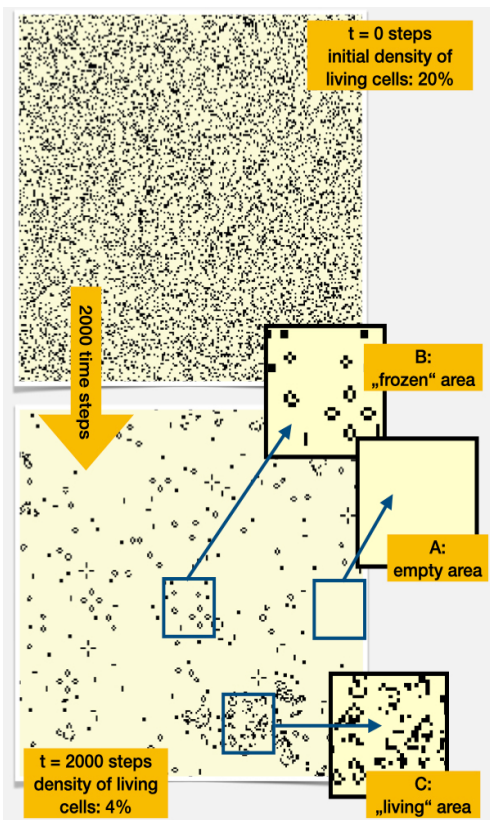

Figure 1: Typical development in the Game of Life.

Figure 1 shows a typical example of a simulation run of the GoL. From an initially randomized distribution of living cells, in this case, 20% of all cells, the pattern formation and self-organization have kicked in and shaped the appearance of the system after the runtime of 2000 time steps that are shown here. The population density has declined dramatically to a population density of approx. 4% in this period of time and the "world" exhibits several areas inhabited by living cells in different types of configurations. The overall landscape can be characterized by 3 archetypical types of areas: Some areas are depleted of living cells (empty areas, A), some contain only fixed patterns (the blocks, the beehives, the loaves, …) or simple in-place oscillating patterns (e.g., the blinkers, …), so the population dynamics came to a standstill ("frozen" areas, B). In parallel, there exist some regions where the pattern formation process is intensively working and producing quickly changing and dynamic patterns ("living" areas, C). Empty areas (A) correspond to Wolfram class I, "frozen" areas to Wolfram class II, while "living" areas best correspond to Wolfram class IV, following the nomenclature given by Stephen Wolfram [17].

Over time, the size ratios between these three types of areas change: They converge towards specific population densities, depending on the random starting density of living cells. This suggests a process driven by a "carrying capacity", reminding the observer of biological density-dependent population dynamics [18].

## 3. EMERGENT POPULATION DENSITIES IN THE GAME OF LIFE

The microscopic rule set of the GoL is purely based on local density information. It is only the fraction of living cells within the population of eight neighboring cells, that determines the fate of every single cell. The position of these living cells plays no role at all, what makes the observed pattern formation already an interesting feature of the system. Given that the microscopic rules are purely dependent on local densities, one might assume that the global density of living cells, concerning their dynamics over time, and concerning potential points of convergence (equilibria, fixed points) should be trivially predictable from these rules.

In order to investigate these dynamics and their points of convergence, a series of experiments was performed. This is necessary to test if those macroscopic properties, that are suggested as candidates for strong emergence in the latter part of this article, really exist in the GoL.

Experiment 1: Parameter sweep of initial densities. The GoL system (size 201 x 201) was initialized with random initial populations of varying densities ranging from 0%, thus $\Omega(0) = 0.0$, to 100% living cells, thus $\Omega(0) = 1.0$, in increments of 1%-wide steps. Each setting was repeated 30 times with the same initial density of living cells, but with a different randomized distribution of those cells. Thus, in total 101 different setups and 3030 simulation runs were performed in experiment 1. Each simulation was run for 5000 time steps. Afterwards, the resulting density of living cells $\Omega(5000)$ was measured, to estimate the LTNPD.

Figure 2 shows the result of the parameter sweep conducted in experiment 1. The dashed lines indicate minima and maxima of observed results, the grey area contains 50% of all results, and the bold black line indicates the median result for any given initial density of living cells.

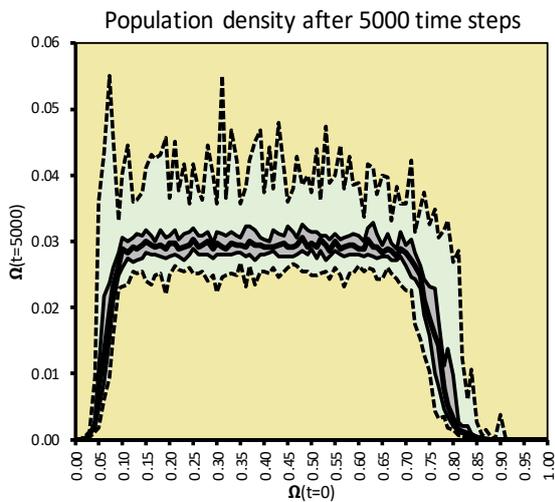

Figure 2: Results of experiment 1. The distribution (quartiles) of the final population sizes at $t=5000$ is shown for each starting density of living cells (N= 30 repetitions per setting).

Figure 2 shows, that there are two significant phase-transitions in the system. There is a critical density of randomly distributed living cells with approx. $\Omega(0) \approx 0.03$, below which the system tends to go fully extinct, thus $\Omega(5000) \approx 0.0$. A less sharp phase-transition occurs for high densities, which start to make the final density decline approx. at $\Omega(0) \approx 0.7$ and above, showing full extinction almost in every case above $\Omega(0) \approx 0.8$. Between the values of $\Omega(0) \approx 0.07$ and $\Omega(0) \approx 0.7$ there seems to be a system regime at work that, in the vast majority of runs, drives the final density to converge towards values around $\Omega(5000) \approx 0.029$. It is noteworthy that the populations seem to converge almost exactly to the critical lower density. Interestingly, the GoL system drives itself to a global population density that is not sustainable if it were a randomized population. These findings correspond well those reported in a study about noisy and asynchronous update procedures of the GoL [19] for some starting conditions. Here, the parameter sweep was conducted in finer steps for the full range of possible initial densities to have a complete view on how the LTNPD depends on these initial conditions.

In order to show that these observations are due to the existence of one or more points of convergence in the emerging population dynamics, a second experiment was conducted to study not only the final population sizes but the dynamics of the populations.

Experiment 2: Here, four different initial conditions were tested with the same GoL system used in experiment 1. These runs were initialized with random initial population densities of living cells $\Omega(0) \in \{0.05, 0.1, 0.15, 0.2\}$. Again, each setting was repeated 30 times with the same initial density but with a different randomized distribution of living cells. Thus, in total 120 experiments were simulated in experiment 2, each one for 5000 time steps.

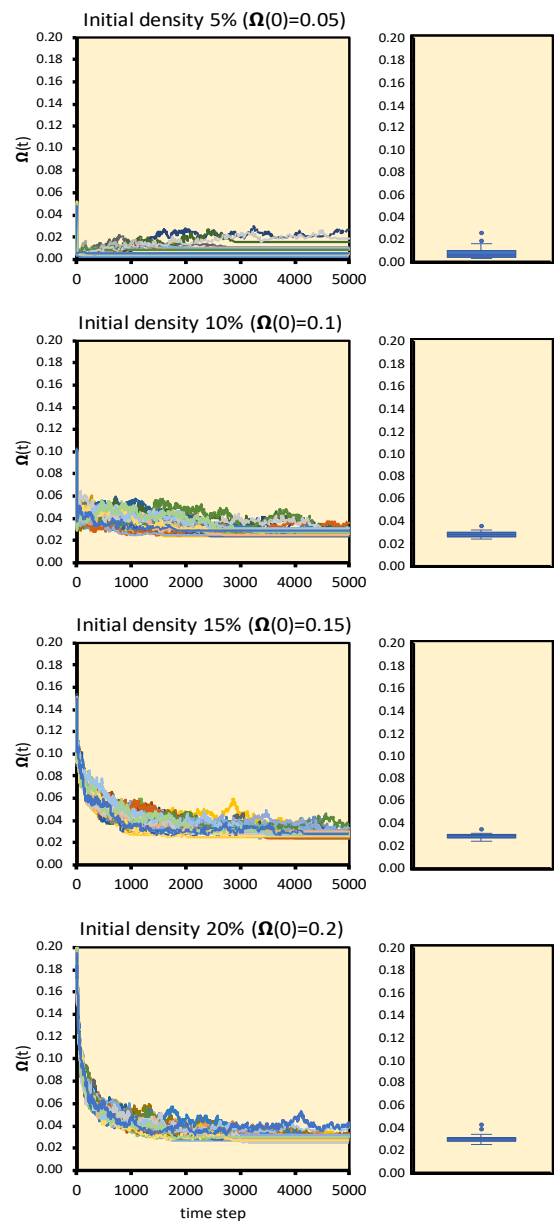

Figure 3: Results of experiment 2. The left column of subfigures shows dynamics of the populations of living cells over time, the right column of subfigures shows the distribution of these populations at the end of the simulations (N= 30 repetitions per setting).

Figure 3 shows that there is clearly a convergence of population dynamics towards the final non-zero population densities reported by experiment 1 for the starting values of $\Omega(0) \in \{0.1, 0.15, 0.2\}$. For the setting of $\Omega(0) = 0.05$, there seems to be a different situation, as a rich variety of outcomes is reached in those runs, none of which tend towards higher population densities than in the other initial settings. However, with such low initial densities, also lower final population densities seem to be possible, indicating that there might be another convergence point (or maybe even more of them) present at these lower densities.

Looking at the individual runs depicted in the left column of Figure 3, it also shows that runs that come to an early standstill (strait horizontal lines in Figure 3), thus reached a configuration with only "frozen" and empty areas, tend to have lower LTNPD than those runs that are still highly dynamic (the wiggling lines in Figure 3), thus are still containing "living" areas: These runs seem also to converge towards an LTNPD residing approximately between $\Omega(t) \approx 0.025$ and $\Omega(t) \approx 0.035$, what is higher than those of the frozen and empty runs. The right column of subgraphs in Figure 3 shows the distribution of the final values of $\Omega(5000)$. These distributions are also numerically given in Table 2.

The results found in both experiments indicate that the GoL system actively approaches an LTNPD, that is a non-zero density equilibrium towards which the population converges in the long term. The same LTNPD was found for a wide range of randomized initial population densities, except for the extremely low and the extremely high initial population densities, which approach either a lower LTNPD or even steer the system towards extinction. The question is now: Can a model predict these sets of behavior qualitatively? Can the observed threshold values for population collapses and the observed values for the non-zero point of convergence be predicted by those models also quantitatively based purely on the known microscopic rules, in order to classify those phenomena as being only weakly emergent?

Table 2: Final distributions of $\Omega(5000)$ in the GoL

|  | $\Omega(5000)$ | | | |
| --- | --- | --- | --- | --- |
|  | Min | Median | IQR | Max |
| $\Omega(0) = 0.05$ | .0024 | .0052 | .0051 | .0257 |
| $\Omega(0) = 0.10$ | .0236 | .0281 | .0036 | .0354 |
| $\Omega(0) = 0.15$ | .0242 | .0289 | .0027 | .0366 |
| $\Omega(0) = 0.20$ | .0246 | .0293 | .0025 | .0421 |

## 4. MACROSCOPIC MODEL

Building a macroscopic model of the observed dynamics from experiment 2, which then could also predict the results of experiment 1, is basically a straightforward task.

The easiest approach would be to take the observed data of experiment 2 and to fit an exponential decay model to the simulation runs with $\Omega(0) \in \{0.1, 0.15, 0.2\}$ and maybe a logarithmic model to the runs with $\Omega(0) = 0.05$. However, there certainly are functions that might support a one-model-fits-all approach, that allows predicting the dynamics for all initial densities $\Omega(0)$. Clearly, one may fit a set of parameters to the observed data to parametrize a statistical post-hoc model, with which the resulting density for any initial density might be predicted. Such an approach would be purely operating on the macroscopic data and not taking any of the microscopic rules into consideration in its model building process. Thus, concerning the observed phenomena, this approach would not inform us about any micro-to-macro causation in the GoL.

Another macroscopic approach, that is resembling the process of life & death in the system, might use a macroscopic model stemming from known ecological literature to describe the population dynamics of living cells. Such a model would likely be a variant of the classical Verhulst model of density-dependent growth [20], as it is described by

$$\frac{\Delta N}{\Delta t} = R \left(1 - \frac{N_t}{C}\right) N_t. \quad (1)$$

In equation 1, the variable $N_t$ is the population of living cells in time step $t$, the parameter $R$ represents a growth rate of the given population, and the parameter $C$ represents a given carrying capacity of the system (e.g., determined by the available space) as the only two constant parameters. I chose the expression as a discrete process on purpose here, as the GoL is an automaton that operates in discrete time steps. The Verhulst model is known to have an unstable point of convergence (equilibrium, fixed point) at $N^* = 0.0$, and another stable point of convergence at $N^{**} = C$, for all above-zero values of $R$, $C$ and $N_o$ [21][22].

Looking closer at the observed results, there is a minimum population density, below which the populations collapse, or get very close to absolute extinction. This might remind biologists on the Allee effect [23], which describes similar dynamics in biological populations. For example, it can become more difficult for animals to find mating partners as population densities become lower. Other examples are that causes for mortality can be avoided better in larger groups than in smaller groups (e.g., predator detection and avoidance) or other forms of social cooperation.

The typical model of the Allee effect is described as

$$\frac{\Delta N}{\Delta t} = R \left(\frac{N_t}{A} - 1\right)\left(1 - \frac{N_t}{C}\right) N_t. \quad (2)$$

Equation 2 holds the same parameters and follows the basic concept of the Verhulst model (equation 1), except for the additional constant parameter $A$, which represents a specific threshold population size, below which the growth rate will become negative. The Allee effect extension of the Verhulst model is known to have a stable point of convergence (equilibrium, fixed point) at $N^* = 0.0$, and another stable point of convergence at $N^{**} = C$, and an unstable equilibrium point at $N^{***} = A$ for all above-zero values of $R$, $C$, $A$ and $N_o$.

It is noteworthy that this model describes the observed phase-transition with low initial population densities in the GoL system, but it does not explicitly model for the phase-transition observed with high initial population densities. However, the negative growth at values of $N_t > C$ increases with increasing distance of $N_t$ from $C$, thus the system is known to be able to show "undershoot" behavior with sufficiently high values of the parameter $R$. This means that also extremely high populations can drive the system towards the extinction point also in this macroscopic model.

In order to test whether the GoL allows a macro-to-macro prediction concerning the focal characteristics observed in the study at hand, another set of simulation runs of the GoL was conducted. Each setting was repeated 30 times and simulations ran for 5000 time steps with randomized initial population densities ranging from 0% to 100% in 5%-wide increments. These additional simulation runs were necessary because the dataset for Figure 2 reported only the final populations and not the course of the dynamics, and the dataset for Figure 3 contained the full dynamics, but only for initial densities up to 20%. The results of these parameter sweeps are shown in subfigures in the left column of Figure 4, they correspond well to the data shown in Figures 2 and 3, thus giving additional confirmation there.

The topmost subgraphs in Figure 4 show the ten first time steps, to make the very specific initial behavior of the system visible, which would be otherwise invisible due to the horizontal scaling of the data. The subgraphs in the middle row show the population density dynamics over the full runtime, and the subgraphs on the bottom row show the initial versus final population densities. As the parameter sweep here is significantly coarser than the one shown in Figure 2, the phase transitions are not captured as precisely as there, however this data suffices for the test if a macroscopic model can be easily fitted to these data.

The subgraphs in the right column of Figure 4 show the corresponding predictions of the model with the best parameter setting found in the model-fitting procedure. For this fitting, the carrying capacity was set to $C = 1172$ living cells in a grid size of 201x201 cells, a value that closely resembles the observed LTNPD in Figures 2 and 3. As the population collapse at the lower density threshold is observed at population densities just slightly below the LTNPD, a value of the Allee effect threshold was set to $A = 1000$ living cells. With these fixed parameter values, the only free parameter value is R, which was systematically varied in a way to minimize the squared differences between the simulation data and the model predictions for all initial densities and over the full course of time. A suitable value for $R$ was found at $R = 0.00075$, as with this value the macroscopic model is capable to predict all focal macroscopic system properties: It captures both population collapses at very high and very low densities, it captures the key population dynamics, and it captures the convergence towards a value close to the observed LTNPD of the GoL. The quality of these predictions can be seen in Figure 4 by comparing the graphs in the left column with those in the right column.

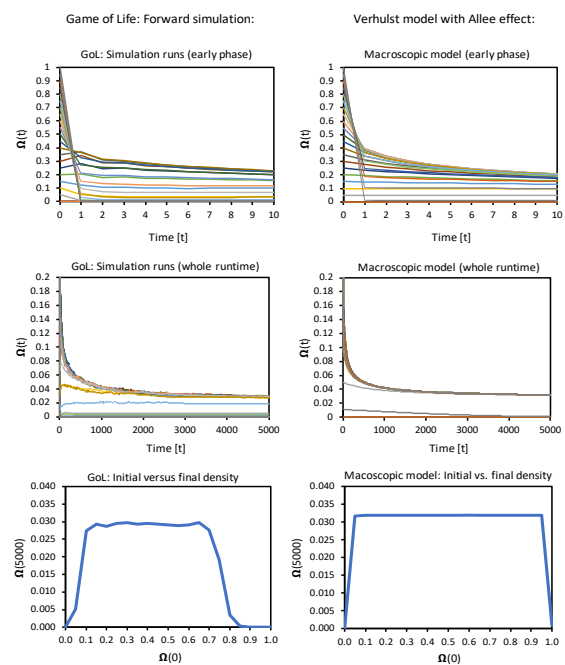

Figure 4: Results of fitting the Verhulst model with Allee effect extension to simulation data from the GoL. Left column of subfigures shows the simulation results of the GoL with various initial population densities, the right column of subfigures shows the corresponding macroscopic model predictions for the same initial conditions.

Although it was easy to fit a plain-vanilla version of the Verhulst model with Allee effect extension (equation 2) to the population data observed in the GoL, this does not tell us why this specific ruleset produces these phenomena. Concerning the (type of) emergence of the focal properties studied in this article, this model-fitting is not helpful per se, because it operates only the macroscopic system layer: The model is constructed with macroscopic thinking, and it was fitted only to macroscopically derived post-hoc data. However, in order to *really understand* the emergent properties of the system, such a modeling approach would only be helpful if we were able to derive the parameter values for $R$, $C$ and $A$ directly and a-priori from the microscopic rule set instead of fitting them post-hoc from observed macroscopic data.

## 5. MICROSCOPIC MODEL

In the following a microscopic model is developed and analyzed to see if these macroscopic model parameters ($R$, $C$, $A$) can be derived directly in an a-priori process from the known rule set without any simulation of the system. Clearly, the value of the parameter $C$, for which both, the Verhulst model and its extension towards the Allee effect are known to converge, should correspond to the LTNPD that we observe in simulations of the GoL system. Can this macroscopically easily observed value of the parameter $C$ be predicted also from the microscopic rules in order to establish the micro-to-macro causation? For this, an approach is chosen that focuses on the feedbacks in the system that are emerging from the specific rules and that can by predicted in an a-priori way directly from these rules.

Figure 5 shows a simple graphical representation of such a microscopic approach, in which arrows show in which direction populations will tend to develop based on individual updates of cells, which are based on their local density of living cells in their Moore neighborhood. It indicates three distinct instances in which the individual cell behavior will not lead to a change in the overall living cell density:

1. Cells that have (on average) 4-6 living neighbors will preferentially die, thus population densities will decrease in consequence, while an (on average) living neighborhood size of 3 living cells can lead to the birth of a new living cell. This will contribute to increasing the population density. This indicates that there is an equilibrium density with conditions where cells have (on average) between 3 to 4 neighbors. This hints towards a stable (regulated) LTNPD of 3.5 living cells (on average) in a population of 8 neighbor grid cells, indicating that there should be expected an LTNPD at a density of 3.5/8 = 0.4375 living cells per grid cells in the system, <u>indicating a stable equilibrium population density of 43.75%.</u>

2. Cells that have (on average) 2 living neighbors should constitute an equilibrium at 2 living cells per neighborhood at 2/8 = 0.25 living cells per grid cells in the system, with no obvious regulation towards this density present in the microscopic mechanics. <u>This indicates an unstable equilibrium population density of 25%.</u>

3. Cells that have (on average) less than 2 living neighbors will preferentially die, thus population densities will decrease towards zero. <u>This indicates a stable equilibrium population density of 0%.</u>

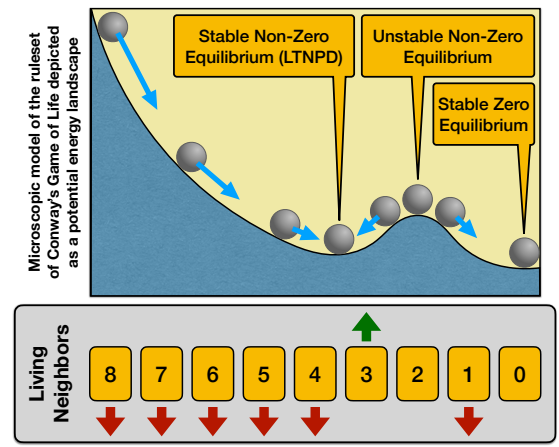

Figure 5: Microscopic model derived from the ruleset of the Game of Life.

These interpretations of the microscopic rules show that a carrying capacity of $C$ of the GoL system must be assumed to be at approx. 43.75%, while it is in fact observed around 2.9% in the simulation experiments (see Figures 2 and 3). Clearly, the micro-to-macro prediction fails by more than one order of magnitude.

Also, concerning the threshold parameter $A$ of the macroscopic model, the values suggested by the microscopic model will be significantly wrong: The unstable tipping point, below which the populations go extinct, is suggested to be at a neighborhood of $2/8 = 0.25$ living cells per grid cells, thus a density of 25%, while it is found in Figure 2 to reside rather around a density of 3%.

Besides that, another macroscopic aspect fails to be deductible even qualitatively: The second phase-transition at high initial densities observed in experiment 1 (see Figure 2). The microscopic model does not provide useful information on why populations with 70% initial population density (on average 5.6 living neighbors per cell) flourishes at a "normal" population density level in the long run, while initial populations with 75% density (on average 6 living neighbors per cell) struggle already significantly and those with 80% density (6.4 living neighbors per cell) are already at the edge of extinction.

## 6. MESOSCOPIC MODEL

The inability of the microscopic model to predict the population dynamics correctly by more than one order of magnitude stems from the fact that the microscopic rules lead to pattern formation which in turn structure the population into areas of different

Wolfram categories (see Figure 1). These distinct area types have different density properties, and we have seen in Figure 2 that density influences further density development.

An obvious choice would be to pursue a mesoscopic model that captures these areal dynamics by depicting the feedbacks that govern their dynamics. Again here, this approach is only helpful if this model can be constructed and parametrized purely by the microscopic processes in an a-priori way from the microscopic system ruleset.

A straightforward approach to such a mesoscopic model would be to subdivide the world of the GoL into an exhaustive set of quadrants which are characterized by the main dynamics that happen inside of them (empty, frozen, living), according to the three area types indicated in Figure 1. The mesoscopic model can then describe the transitions of quadrants from one type to the other. For example, following a System Dynamics model building approach [24], one can implement a system of ordinary differential or difference equations (ODEs or OΔEs) along with the concept of a Stock&Flow modeling approach [25], as is shown in Figure 6.

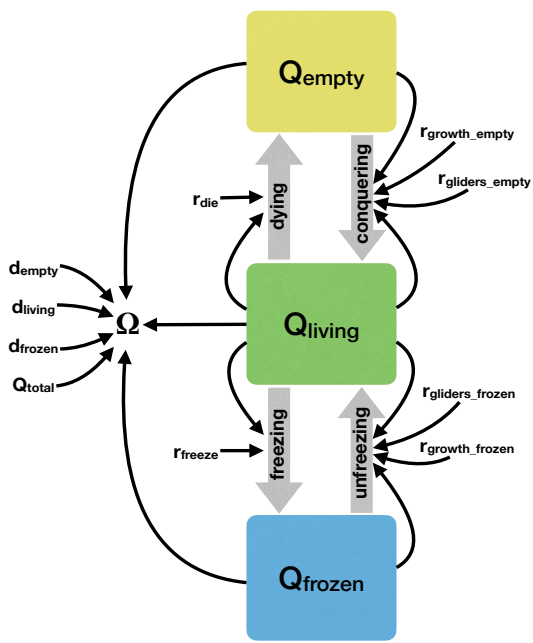

Figure 6: Mesoscopic Stock&Flow model of the population dynamics in the Game of Life.

The colored boxes in Figure 6 indicate the key system variables in such a model: Quantities of quadrants in specific states. Grey thick arrows indicate transitions of quadrants from one state to another. Black thin arrows indicate causal dependencies of system components. Text elements without boxes indicate all constant parameters and the only other system variable $\Omega(t)$. This focal macroscopic variable can be predicted based on the ratios of these three quadrant populations, based on the population densities that these quadrants exhibit ($d_{empty}$, $d_{living}$, $d_{frozen}$) and based on the absolute number of quadrants. Four distinct flows describe transitions of quadrants from one type to another, ensuring conservation of mass in the system. Such flows would be modeled as follows:

- A fraction of the "living" quadrants can either freeze or get empty, at rates set by the two coefficients $r_{freeze}$ and $r_{die}$. This process is independent from the existing population of frozen and empty quadrants, as it is indicated in Figure 6 with two thin black arrows pointing at the left two flows: These flows are influenced by $Q_{living}(t)$, but not by $Q_{empty}(t)$ and $Q_{frozen}(t)$.

- A fraction of the "frozen" quadrants can become living again (unfreezing). This can only happen due to activities that happen in "living" quadrants. This codependency will be best expressed with a mass-action law term $Q_{frozen}(t) \cdot Q_{living}(t)$, multiplied with the sum of two coefficients: The coefficient $r_{growth\_frozen}$ describes the rate at which a population of living cells grows beyond its quadrant borders and re-triggers the "living" process in a neighboring quadrant. Another coefficient $r_{gliders\_frozen}$ describes the rate at which "living" quadrants produce gliders or other moving structures that enter a "frozen" quadrant and re-triggers the "living" processes there.

- Analogously, a fraction of the empty quadrants can be resettled with life by two similar processes: Either they are conquered by a neighboring "living" quadrant through growth ($r_{growth\_empty}$) or they are bootstrapped by two gliders, potentially originating from distant "living" quadrants ($r_{gliders\_empty}$). Both processes will be best modeled by a mass-action law term $Q_{empty}(t) \cdot Q_{living}(t)$ as it requires both, "living" quadrants and empty quadrants that interact.

This model, and especially the usage of the mass-action law approach to describe how patches (quadrants) inhabited with life "conquer" uninhabited ones, is closely following the classical Levins model, that is used for predicting biological settlement and spreading processes [26].

In contrast to the Levins model, which considered only living and dead patches, the mesoscopic GoL model here is extended by considering also a "frozen" habitat state. However, the basic modelling structures and concepts are similar for both non-"living" quadrants, as they can only get alive by interaction with existing "living" quadrants. Consequently, the model will have four different equilibria: Three unstable equilibria at $[Q_{empty}(t) = Q_{total}, Q_{living}(t) = 0, Q_{frozen}(t) = 0]$, at $[Q_{empty}(t) = 0, Q_{living}(t) = 0, Q_{frozen}(t) = Q_{total}]$ and also at $[Q_{empty}(t) > 0, Q_{living}(t) = 0, Q_{frozen}(t) > 0]$. With no-zero flow coefficients there will be also a stable equilibrium at a quadrant population of $[Q_{empty}(t) > 0, Q_{living}(t) > 0, Q_{frozen}(t) > 0]$.

Even though the mesoscopic model is more complicated than the microscopic one, it will fail to predict the two phase-transitions completely, as this mesoscopic model structure can neither predict the Allee effect type behavior nor the population collapse at high densities. For implementing these, the model would need to be extended with additional terms, again carrying the parameters $A$ and $C$, which are also not correctly derivable from the microscopic rules, as was shown already before.

This mesoscopic model has ten parameters that need numeric values: (1) the six abovementioned coefficients (the six "$r$"-parameters in Figure 6) for the four flows in the Stock&Flow model, (2) the three population densities ($d_{empty}$, $d_{living}$, $d_{frozen}$) that the quadrant types will exhibit, and (3) the total number of quadrants $Q_{total}$.

In order to be useful as a micro-to-macro model, all the parameter values need to be determined from the microscopic mechanisms. Besides the trivial determination of $Q_{total}$, only two density parameters ($d_{empty}$ and $d_{living}$) can be derived from microscopic thinking: The value of $d_{empty}$ must be *0.0* and that $d_{living}$ should be around *0.4375*, as this density was deducted from the microscopic model already for those areas where the Game of Life is actively processing its own state in a highly dynamic manner. The remaining density estimator $d_{frozen}$ and all six process coefficients seem to be impossible to be deducted from a microscopically informed deduction. The densities that are observed for such areas (see Figure 1) are much lower than the so-called "maximum density still life" (MDSL) that "frozen" areas can have, which is found to be at 57.1% [27] and also lower than the 25% that are suggested as the unstable equilibrium with (on average) 2 out of 8 neighboring cells being "living" by the microscopic model depicted in Figure 5. Thus, there is no knowledge stemming from a microscopic ruleset interpretation that would suggest a useful prediction of the observed density towards which "frozen" areas emergently converge during the time progression of the GoL simulations.

Deriving the flow coefficients from the microscopic rules is the next unsolved problem: Neither do I see a way to derive the rate of glider production, the rate of glider-to-glider collisions, and the rate of glider-to-frozen-structure collision, nor can I derive the rates of freezing, growth, and death directly from the microscopic rules, as it would be needed.

The bottom-up model-building task feels more and more like going down a rabbit hole: Instead of one problem, as it was the case with the purely microscopic model depicted in Figure 5, there are now seven separate problems to be solved in the model depicted in Figure 6. To derive each needed parameter value microscopically, a specific sub-model is required. These seven sub-models will likely require building even more sub-sub-models and so forth. This can be an indicator that it is in-principle impossible to make a consistent micro-meso-macro model that can predict the population dynamics and the LTNPD in the GoL what would mean that this emergent phenomenon stems from strong emergence.

Besides these problematic parametrizations, there are other problems that arise with the mesoscopic model: The concepts that are applied in model building are not stemming from the microscopic knowledge. For example, the fact that "frozen" areas require interaction with "living" areas to become alive again can hardly be a-priori derived from studying the simple microscopic rules. It is knowledge from studying simulation runs of the system, thus this model building already incorporates after-the-fact knowledge rather than only a-priori knowledge.

## 7. STUDYING OTHER VARIANTS

This section considers alternative settings and questions whether the beforementioned results are just specific artifacts of the initial conditions or the specific ruleset. So far, the GoL has always been studied with randomly distributed distributions of living cells in the study presented here. It is known that a simple structure, called the "R-pentomino" configuration [28], shows impressive growth by starting with only 5 adjacent living cells in an otherwise empty world. This allows to test if the GoL approaches the observed LTNPD also from very low starting points concerning its initial population density.

Additionally, it is investigated in the following if other cellular automata, that are very closely related to the ruleset of the GoL, exhibit a similar failure of correctly predicting their LTNPD directly with a micro-to-macro model, or not.

### 7.1. STUDYING DYNAMICS STARTED FROM SPARSELY DISTRIBUTED R-PENTOMINO POPULATIONS

For this analysis, the same settings were used as in the previous experiments. The only difference is that this time ten R-pentomino configurations were initially placed at random positions on the grid, all other cells were set to the "dead" state. This corresponds to an initial density of $\Omega(0) = 0.00124$ (0.124% density). Figure 7 shows one instance of the R-pentomino configuration, the population dynamics of 30 simulation runs, and the final distribution of the approached LTNPD. It is clearly visible, that even from this low initial starting point, a population density is actively approached that is very similar to the results obtained in the previous experiments. The observed median density, that the system converges to, was found to be at $\Omega(5000) \approx$ 0.024 and the observed IQR was 0.004. Like the previous experiments, those runs that in the end still had "living" areas achieved larger final populations than those that were already fully composed from "frozen" and empty areas. This indicates that the observations made here, are not artifacts of randomized initial distributions, but that density-dependent growth dynamics hold also for very low starting populations if they are configured in a "viable" way (e.g., with a few R-pentominoes).

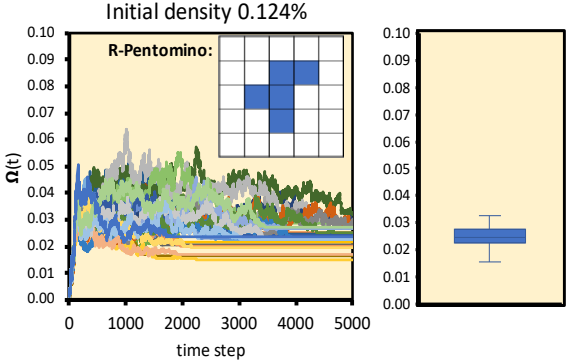

Figure 7: Population dynamics and the distribution of the final population after initialization with a low density of R-pentominoes.

## 7.2. STUDYING CLOSELY RELATED CELLULAR AUTOMATA

In order to investigate how "special" the LTNPD of the GoL is concerning its type of emergence, several cellular automata were tested with slightly adapted rulesets. The key research question of these experiments is: Do micro-to-macro model predictions also fail the LTNPD that these cellular automata converge to?

### 7.2.1. VARIANT A: REPRODUCTION ALSO WITH MORE LIVING NEIGHBORS

The rules of this cellular automaton differ from the GoL by the fact, that a dead cell with four living neighbors will become alive, in addition to the birth with three living neighbors inherited from the GoL. Table 3 indicates the new rule with a solid frame, while the rules inherited from the GoL are indicated by dashed frames.

Table 3: Ruleset of the variant A cellular automaton

| if the focal cell *at (x,y)* is dead at time step *t*: | | | | | | | | | |
|---|---|---|---|---|---|---|---|---|---|
| N(x, y, t) | 0 | 1 | 2 | 3 | 4 | 5 | 6 | 7 | 8 |
| S(x, y, t) | 0 | 0 | 0 | 0 | 0 | 0 | 0 | 0 | 0 |
| S(x, y, t+1) | 0 | 0 | 0 | 1 | 1 | 0 | 0 | 0 | 0 |
| if the focal cell *at (x,y)* is alive at time step *t*: | | | | | | | | | |
| N(x, y, t) | 0 | 1 | 2 | 3 | 4 | 5 | 6 | 7 | 8 |
| S(x, y, t) | 1 | 1 | 1 | 1 | 1 | 1 | 1 | 1 | 1 |
| S(x, y, t+1) | 0 | 0 | 1 | 1 | 0 | 0 | 0 | 0 | 0 |

Figure 8 shows that this cellular automaton converges to a fully "frozen" maze-like pattern. The microscopic model for this ruleset, analog to the one shown in Figure 5 for the GOL, predicts the LTNPD between four and five living neighbors, pointing at a value of $\Omega(\infty) = \frac{4.5}{8} = 0.5625$. This suggests a convergence towards a population density of approx. 56%. In 30 repetitions of 5000 step long simulations of this cellular automaton, the LTNPD was found to be on a median value of $\Omega(5000) \approx 0.507$, or approx. 51%, which is very close to the value predicted with the purely microscopic model.

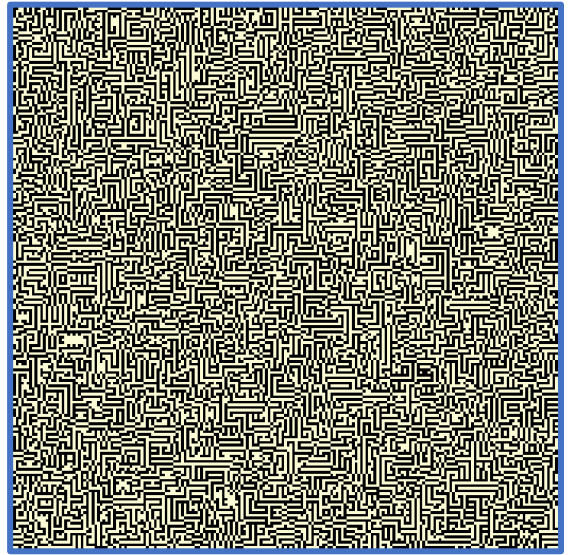

Figure 8: Typical "frozen" pattern produced by the variant A cellular automaton.

### 7.2.2. VARIANT B: REPRODUCTION WITH FEWER LIVING NEIGHBORS & SURVIVAL IN A NARROWER RANGE

The rules of this cellular automaton differ from the GoL by the fact, that a dead cell with two living neighbors will become alive, and not with three living neighbors, as it is the case in the GoL. In addition, cells with three living neighbors will die, while such cells will stay alive in the GoL. Table 4 indicates the new rules with solid frames, while the survival rule inherited from the GoL is indicated by a dashed frame.

Table 4: Ruleset of the variant B cellular automaton

| if the focal cell *at (x,y)* is dead at time step *t*: | | | | | | | | | |
|---|---|---|---|---|---|---|---|---|---|
| N(x, y, t) | 0 | 1 | 2 | 3 | 4 | 5 | 6 | 7 | 8 |
| S(x, y, t) | 0 | 0 | 0 | 0 | 0 | 0 | 0 | 0 | 0 |
| S(x, y, t+1) | 0 | 0 | 1 | 0 | 0 | 0 | 0 | 0 | 0 |
| if the focal cell *at (x,y)* is alive at time step *t*: | | | | | | | | | |
| N(x, y, t) | 0 | 1 | 2 | 3 | 4 | 5 | 6 | 7 | 8 |
| S(x, y, t) | 1 | 1 | 1 | 1 | 1 | 1 | 1 | 1 | 1 |
| S(x, y, t+1) | 0 | 0 | 1 | 0 | 0 | 0 | 0 | 0 | 0 |

Figure 9 shows that this cellular automaton converges to a highly dynamic snow-storm-like pattern with occasional short-termed aggregation of living cells. The microscopic model, analog to the one shown in Figure 5 for the GOL, predicts the LTNPD between two and three living neighboring cells, pointing at a value of $\Omega(\infty) = \frac{2.5}{8} = 0.3125$. This suggests a convergence towards a population density of approx. 31%. In 30 repetitions of 5000 step long simulations of this cellular automaton, the LTNPD was found to be on a median value of $\Omega(5000) \approx 0.271$, or a density of approx. 27%, which is again very close to the value predicted by the purely microscopic model.

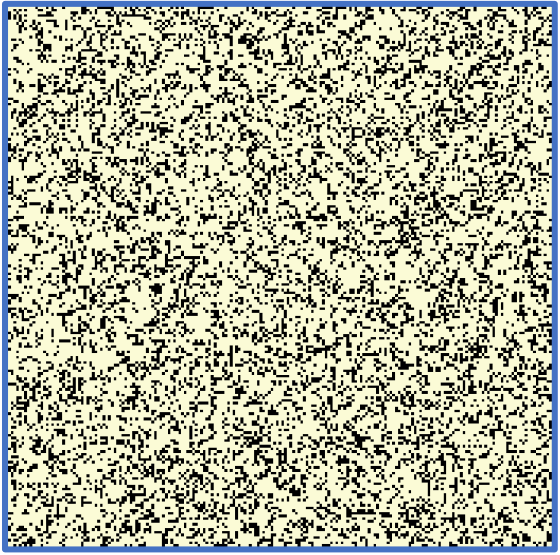

Figure 9: Typical (highly dynamic) pattern produced by the variant B cellular automaton.

### 7.2.3. VARIANT C: REPRODUCTION WITH MORE LIVING NEIGHBORS & WIDER SURVIVAL RANGE

The rules of this cellular automaton differ from the GoL by the fact, that a dead cell with four living neighbors will become alive, and not with three living neighbors, as it is the case in the GoL. Living cells with one to four living neighbors will survive, while in the GoL only cells with two or three living neighbors stay alive. Table 5 indicates the new rules with solid frames, while the survival rules inherited from the GoL are indicated by dashed frames.

Table 5: Ruleset of the variant C cellular automaton

| if the focal cell *at (x,y)* is dead at time step *t*: | | | | | | | | | |
|---|---|---|---|---|---|---|---|---|---|
| N(x, y, t) | 0 | 1 | 2 | 3 | 4 | 5 | 6 | 7 | 8 |
| S(x, y, t) | 0 | 0 | 0 | 0 | 0 | 0 | 0 | 0 | 0 |
| S(x, y, t+1) | 0 | 0 | 0 | 0 | 1 | 0 | 0 | 0 | 0 |
| if the focal cell *at (x,y)* is alive at time step *t*: | | | | | | | | | |
| N(x, y, t) | 0 | 1 | 2 | 3 | 4 | 5 | 6 | 7 | 8 |
| S(x, y, t) | 1 | 1 | 1 | 1 | 1 | 1 | 1 | 1 | 1 |
| S(x, y, t+1) | 0 | 1 | 1 | 1 | 1 | 0 | 0 | 0 | 0 |

Figure 10 shows that this cellular automaton exhibits an intriguing behavior: It is highly dynamic first, but then zones "freeze" into patterns of horizontal or vertical stripe patterns, between which dendritic-like empty patterns form, that show high activity along their edges. Slowly, the frozen areas grow at the cost of the empty dendritic areas. In the end, reminiscent traces of these dendritic areas remain as empty islands in the end, and finally only frozen and empty areas remain. The microscopic model, analog to the one shown in Figure 5 for the GOL, predicts the LTNPD between four and five living neighboring cells, pointing at a value of $\Omega(\infty) = \frac{4.5}{8} = 0.5625$. This suggests a convergence towards a population density of approx. 56%. In 30 repetitions of 5000 step long simulations of this cellular automaton, the LTNPD was found to be on a median value of $\Omega(5000) \approx 0.4804$, or approx. 48%, which is again rather close to the value predicted by the purely microscopic model.

It is noteworthy with the results obtained with variant C, that compared to the other two variants, this automaton exhibits a more interesting and more complex transient towards its final configuration. Thus, the fact that it also shows the greatest deviation from the microscopic predictions amongst all variants, hints towards a potential tendency that interesting automata may generally show higher error in their micro-to-macro predictions, turning this error into a telltale sign of potential interesting (stronger) emergence.

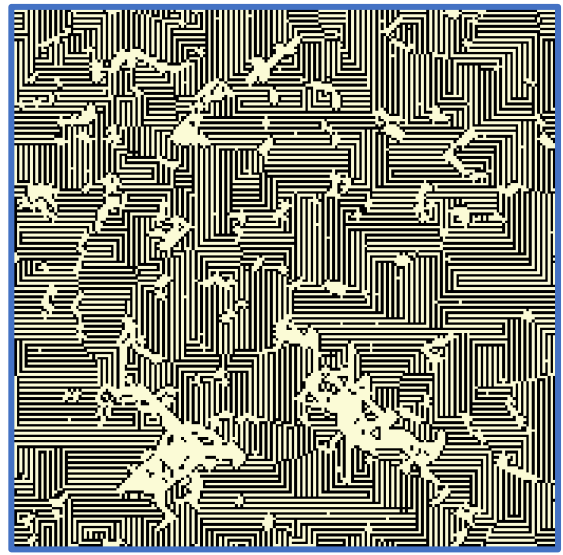

Figure 10: Typical pattern produced by the variant C cellular automaton.

## 8. DISCUSSION & CONCLUSION

In the 50 years since its first publication, Conway's Game of Life has attracted massive scientific interest. For example, the book at the 40th anniversary of the GoL, edited by Andrew

Adamatzky, is showcasing an exceptional richness of diversity of research that is conducted on and around the GoL [15]: Research is done on the classic GoL itself, for example by studying up to which extent the GoL can endure erroneous and asynchronous update regimes [19]. Additionally, also alternative variants are often created and studied. Such GoL variants have alternative rulesets [29], or they operate with continuous states [30], to mention just a few out of many examples reported in literature. Recently, the impressive "Lenia" system showed that its "Orbium" creatures, which are glider-like entities that move through a continuous space, can be derived microscopically from the rules of the GoL [31], turning Lenia into an advanced continuous-space variant of the GoL. Most prominently, the GoL was found to be Turing-complete in itself, and allowing to construct exceptional structures, e.g., building a universal Turing machine inside of the GoL [32]. Meanwhile, the search for a 3D equivalent of the GoL is an ongoing challenge [33][34].

The study of Mark Bedau suggested the discrimination between weak and strong emergence, and it characterizes the GoL solely by its pattern formation process and pentomino growth. It (correctly) identifies these processes as being of the weak type of emergence [10]. This presentation and its arguments somehow leave the reader with the impression that there are no phenomena of strong emergence in the GoL, although this is not explicitly stated, but strong emergence in general is doubted there.

In contrast to this impression, the study presented here demonstrates that several properties of the GoL are in fact good candidates for strong emergence: The two thresholds at which randomized populations collapse, the density towards which "frozen" areas develop in the GoL and, most prominently, the long-term non-zero population density, that the global system converges to from most of its starting states.

The reason to suggest them as cases of strong emergent are: Neither was the straight-forward, purely a-priori informed, microscopic model, nor was a straight-forward, purely a-priori informed, mesoscopic model found to be capable of delivering even somehow useful hints towards the numerical values of these emergent system properties. In contrast to that, a well-known and well suitable macro-to-macro model exists in ecological modelling literature. It was demonstrated that even the very basic plain-vanilla variant of this macroscopic population model is already very well capable to capture all investigated macroscopic behaviors of the GoL, as it is described in the data shown in Figures 2, 3 and 7. The striking point here is: While this can obviously be easily done in a post-hoc way based on macroscopically derived data, gathering knowledge through direct micro-to-macro causation fails to work sufficiently also here. The known microscopic rules cannot a-priori inform about suitable parameter values for the macroscopic model. And values, that lead to good predictions of the macroscopic model after post-hoc model fitting, are not those value that are suggested by the analysis of the microscopic model. Thus, as the consequence of all these observations reported here across various model building attempts, it is suggested that the observed difficulties, or even inabilities, in the micro-to-macro model building might be an indicator that these investigated global system properties of the GoL are indeed of the strong emergence type.

In reflection of these observations, the study presented here develops the suggestion, but does of course not deliver any definitive proof, that these macroscopic properties of the GoL system could be of the strong emergence type. This suggestion is not only supported by the inability to find a direct micro-to-macro model of sufficient prediction quality, but also by the "going-down-the-rabbit-hole"-type of development in the search for an appropriate mesoscopic model. Usually, the number of unknowns (parameters needed) increase the deeper the model building goes downwards from macro to micro in a top-down way, and it decreases in the bottom-up model building, as macroscopic models are more abstract than microscopic ones. Surprisingly, in the model building for the GoL, the opposite tendency is found. While all needed parameters for the microscopic model were easily derivable from the ruleset and only the resulting predictions were wrong, the mesoscopic model already failed in the parametrization for seven crucially needed system parameters. Trying to parametrize those coefficients will require even more specific sub-models, in turn causing even more parametrization problems.

If the discussed system properties are in fact strongly emergent, this basically means that the category of strong emergence does not require any "magic" or "metaphysics", as it is sometimes called. In contrast, it can appear in a complex causation-driven system, even in a deterministic one like the GoL.

The irreducibility of the global macroscopic states to purely local microscopic dynamics can be caused by micro-macro-micro feedback. For example, local densities are driven by the known microscopic rules (see Figure 5) and local densities affect the behavioral state of local areas, characterizable as "empty", "frozen" or "living". However, the microscopic mechanics find significantly different input patterns in such local area types, thus they produce different follow-up dynamics as their functional output. This then, in turn, affects the dynamics found on higher system layers, e.g., the growth, the decay, and the rate of interaction of such local areas, ultimately determining the proportions of larger areas (quadrants) of specific kinds, yet even the spatial distribution of these quadrant types.

These mesoscopic interactions can evolve their own higher-level mechanics, as it is depicted in Figure 6, which then can produce ultimately the macroscopic behavior, as it is shown in Figure 2 and Figure 3.

As higher-level components change dynamically, for example by quadrants altering their type, this feeds back downwards onto the input configurations that their local microscopic mechanics operate upon: The microscopic rules operate on local densities, and different types of quadrants converge towards different levels of densities. Thus, there is clearly micro-meso-micro feedback within the system, that is ultimately reflected also on the macroscopic system level, as several of such mesoscopic models can be nested inside of each other, all representing quadrants of increasing sizes. Thus, the system can be seen as a micro-meso$^*$-macro-meso$^*$-micro feedback system, having potentially an infinite number of mesoscopic system layers for an infinite grid size.

Even with finite grid sizes of the GoL system, the mesoscopic model depicted in Figure 6 can also be implemented recursively, by nested models of smaller quadrants contained inside of larger quadrants. This way, several mesoscopic layers of the system can be represented mathematically. However, such a many-mesoscopic-layers approach can neither go fully down to the very bottom, to the local microscopic layer, nor can it fully go up to the top, to the global macroscopic layer, for very different reasons:

At the bottom of the system layers, the Stock&Flow model depicted in Figure 6 cannot be applied with quadrants of 3x3 cells in a meaningful way, because (a) the microscopic model depicted in Figure 5 is needed to describe this system layer to incorporate the ruleset of the GoL into the overall model, and (b) such a small quadrant size will likely not allow making the necessary discrimination between „frozen" and „living" quadrants. Very likely such a modeling approach will break already before going fully down to the microscopic level, as for example also 4x4 or 5x5 quadrants will be too small to determine the quadrant type.

At the top of the system layers, the Stock&Flow model depicted in Figure 6 will also break when quadrant sizes get too large. For example, at the topmost macroscopic level, such a Stock&Flow model would then handle only one single quadrant, which changes its type over time. In such a configuration, the flows would operate in a very special and extreme way, as they only can have the values 0 or 1, whereby only one of the four flows and only one of the three stocks can hold a value of 1 at each time step $t$. In such a binary-values-only model, the core concepts of the mesoscopic model will break: For example, the mass-action-law in the modeling of two flows will break, because only one stock will have a non-zero value, thus all associated flows will always remain at zero and all system dynamics will come to a standstill at $t = 1$, the latest. Again here, such a modeling approach will already break before going up to the topmost macroscopic layer, it will already not work in a meaningful way with only a few quadrants depicting the whole system quadrants representing the total grid space.

In general, micro-macro-micro feedback loops impose upwards and downwards causation and can be interpreted as an indicator for strong emergence. The feedback loop has the pattern formation process (a weakly emergent phenomenon) as a component of the loop. This makes it questionable if such a distinction is useful at all. Given that the contribution of the weakly emergent component to the whole feedback loop can vary, the weight of the weak emergence inside the strong emergence loop is variable. This suggests that it is not a dichotomy, it will be rather a spectrum on which the strength of the emergence may fall, potentially correlated with the magnitude of error with which micro-to-macro predictions fail.

Such micro-macro-micro loops are also present in the studied variants of cellular automata. For example, the variant A and C show also a very clear pattern formation process and, in their transient, all three types of areas emerge, like in the GoL. Also, similarly to the GoL, these variants end up in a state with only frozen and empty areas. Despite all these microscopic and macroscopic similarities, their micro-to-macro predictions fit quite well to the measured macroscopic outcomes of their simulation runs, in very contrast to the GoL, which fails there by a full order of magnitude. This indicates that there is a hidden game changer somewhere in the rules of the GoL, which has not yet been identified. Neither the very early "freezing" variants A and C, nor the always highly dynamic ("living") variant B have any problems delivering straightforward and reliable micro-to-macro predictions. What is it in the GoL ruleset that prevents even a somehow reliable micro-to-macro prediction? Given that the rules of the variants are just slight mutations of the GoL rules, this is a fascinating difference in a key property of the system: micro-to-macro predictability.

Ultimately, all these considerations suggest, that the root cause for the inability of micro-to-macro predictions is not the sheer existence of a micro-macro-micro feedback, but it is rather the fact that a weakly emergent functionality (pattern formation) drives this micro-macro-micro feedback system in a way that no reliable model can be made, because of the processes that happen on the mesoscopic system layers. This suggests that exactly this model-building inability will be the suitable distinguishing feature for identifying strong emergence.

It is surprising, that the system properties of the GoL, that are found to be promising candidates for strong emergence, are by far not as spectacular as the pattern formation, which is (only) weakly emergent.

In other words, the most intriguing system properties are sort of hidden in plain sight here, and they seem to have been overlooked and, in consequence, also understudied massively in the past. This is likely due to the impressiveness of the pattern formation process and its follow-up consequences, which attract research more easily and thus to a higher extent.

For example, when looking at how the GoL inspired the striving domains of Artificial Intelligence (AI) and Machine Learning (ML) research, a similar picture is found: It is striking that there is much effort to develop an AI that is capable to learn the microscopic rules from training runs based on macroscopically derived data, e.g., known patterns to be formed [35]. In contrast to that, there seems to be no literature yet that tries to solve the much more difficult other way: This would be to have an AI predicting the hard-to-predict macroscopic outcomes of cellular automata from purely the microscopic input (the ruleset) without iteratively simulating the system over time. Such a trained model would then need to be able to make reliable micro-macro predictions for the GoL and for the variants that are presented here. If these predictions then hold for a sufficient variety of starting conditions across all rulesets, then truly this AI will have cracked the prediction problem and identified the system properties that are potentially of strong emergence, as being cases of only weak emergence.

Such an AI-driven micro-to-macro model building method would be also useful to serve as an automated emergence detector and classifier: First, simulations of complex dynamic systems can be analyzed with a pattern-recognizing AI. If it detects a pattern formation process, this system can be flagged as showing emergent properties. Then a model-building AI can try to learn a predictive model that correctly predicts this system's macroscopic behavior without simulation, straight from the ruleset, together with correct predictions of altered rulesets. If correct predictions are learned by this AI for all systems, then this system can be classified to exhibit only weakly emergent properties. But if the AI's micro-to-macro model learning fails for the focal system, but works well for the other variants, then this system can automatically be flagged as a very promising candidate for strong emergence. Analogously, with such a technology one could concentrate on a single specific focal system and automatically browse through various macroscopic system properties and classify them concerning their emergence type.

The need for such computational AI algorithms to autonomously find interesting emerging properties is useful in massive parameter sweeps through microscopic system parameters, or in evolutionary computation algorithms, that also alter microscopic parameters in high amounts of instances. An example for the first case would be the massive parameter sweep that was conducted with the complex (GoL-like) primordial particle system [36], an example for the second case are the evolutionary approaches conducted with a complex system called "swarm chemistry", where automatic detection of interesting system behaviors would also come very handy, given its "open-ended" evolutionary approach towards finding interesting and novel complex systems [37].

The questions around emergence and its reducibility of phenomena to micro-to-macro models arises with all complex adaptive systems [38], thus they are tied into research questions posed at the most intriguing frontiers of science, where researchers are studying the universe, the phenomenon of life in general, the human brain and the societies and their cultural properties, that are emerging amongst interacting people. While the GoL is just a very simple toy model of a complex dynamic system, studying its hard-to-crack properties may help us to make significant progress in all other domains that have to tackle emergent phenomena in complex systems.

The fact, that the toughest-to-understand properties are appearing to be not spectacular at all and seem to have been hiding in plain sight for a long time, while weak forms of emergence stand out spectacularly and attract much more research, makes me wonder: How many other, yet undetected, instances of such properties are out there, in the GoL, but also in other complex adaptive systems?

## ACKNOWLEDGEMENTS

This work was supported by the Field of Excellence COLIBRI (Complexity of Life in Basic Research and Innovation) at the University of Graz.